\documentstyle[graphicx,float]{article}

\begin{document}
\title{Atomic Kapitza-Dirac effect with quadrupole transitions}
\date{}
\author{Pedro Sancho \\ Centro de L\'aseres Pulsados, CLPU \\ Parque Científico, 37085 Villamayor, Salamanca, Spain}
\maketitle
\begin{abstract}
Interactions between atoms and light fields are usually described in
the electric-dipole approximation. We show that electric-quadrupole
terms are important in the Kapitza-Dirac arrangement for light
gratings on resonance with a quadrupole atomic transition. We derive
the diffraction patterns, which in some cases are experimentally verifiable with
the same techniques used with dipole transitions.
\end{abstract}

\vspace{3mm}


\section{Introduction}

The interaction between atomic systems and light fields is usually
described in the electric-dipole approximation. However, there are
some scenarios where one must go beyond that approximation. In this
paper we want analyze a situation where the electric-quadrupole
term becomes important in the Kapitza-Dirac arrangement.

In the Kapitza-Dirac effect a beam of atoms or electrons is
diffracted by a standing light wave, usually a laser \cite{KD, Bat,
Pri, Ba2, Ba3}. Recently, the theory of the effect has been extended to
two-particle systems \cite{Sa1,Sa2}. In the case of atoms, one
chooses the wavelength of the optical grating in the proximity of an
atomic transition, in order to enhance the strength of the
interaction. Most times this is a dipole transition, and the dipole
transition matrix element is the only relevant in the problem.

An interesting situation emerges when, instead, the optical wavelength
is close to a quadrupole transition. Then the quadrupole transition
matrix element becomes dominant. A similar situation was considered
in \cite{Moi} for the trapping of cold atoms in optical lattices.

We shall analyze in detail the scenario in order to estimate when
the quadrupole-type diffraction is experimentally verifiable. We
shall evaluate the diffraction patterns, which are similar to those
derived in the dipole approximation, but with different values of
the parameters. If we consider a two-mode standing light wave
containing simultaneously the dipole- and quadrupole-resonant
frequencies we can observe a richer behaviour.

\section{Quadrupole transitions}

As signaled in the Introduction, if we choose the wavelength of the
standing light wave close to a quadrupole transition, the quadrupole
term will be dominant. This is the case when the light detuning with
respect to the atomic frequency couples a s-type ground atomic state
with an excited d-type one, instead of a p-type one as it is the
case in dipole transitions. In Calcium, for instance, the lowest
excitation is from this type.

In order to analyze this scenario we must consider the interaction
ruling the transition. We closely follow the presentation in
\cite{Moi}. A beam of atoms interacts with a linearly polarized
laser. We denote by $x$ the propagation direction of light and by
$z$ the oscillation one. The relevant interaction in the problem is
given by the potential
\begin{equation}
V=\frac{e}{mc}A(X,x,t) \hat{p}_z
\end{equation}
with $m$ the mass of the electron, $\hat{p}_z$ the z-component of
its momentum operator, $X$ the coordinate of the center of mass of
the atomic system and $x$ the relative one \cite{Moi}. Being
essentially the Kapitza-Dirac effect an one-dimensional problem, we
restrict our considerations to that longitudinal variable (parallel
to the grating). The transversal or perpendicular variables are
important determining the duration of the interaction, but do not
play any additional relevant role in the diffraction pattern.

The electromagnetic potential at the position of the electron
suffering the transition in the atom can be expressed as $
A(x_{e},t)=A(X+x,t)$, In the dipole approximation the dependence on
the relative coordinate can be neglected and we have
$A(x_e,t)\approx A(X,t)$. In contrast, when the quadrupole term is
taken into account we must consider the next order in the multipole
expansion:
\begin{equation}
A(x_e,t)\approx A(X,t) + x \left( \frac{\partial A}{\partial
X} \right) (X,t)
\end{equation}
If we choose, as usual, for the standing light $A(X) =A_0 \cos (k_L
X)\cos (\omega _L t)$ with $k_L$ and $\omega _L$ the wave vector and
frequency of the laser beam, the contribution of the quadrupole term
will have the form $-A_0 k_L \sin (k_L X)\cos (\omega _L t)$.

With this expansion the interaction potential decomposes as
\begin{equation}
V = \frac{e}{mc}A(X,t) \hat{p}_z + \frac{e}{mc} x \left( \frac{\partial
A}{\partial X} \right) (X,t)   \hat{p}_z =V_D+V_Q
\label{eq:hap}
\end{equation}

The optical potential \cite{Ada,Moi} associated with this interaction is
\begin{equation}
V_{op}=\frac{|<e|V|g>|^2}{\hbar \Delta}
\end{equation}
with $e$ and $g$ the excited and ground atomic states, and $\Delta $
the detuning of this transition frequency with the laser one.

Oscillating the light variables with a periodicity much smaller than
the time scale of the atomic center of mass we must average over
that periodicity, $\tilde{V}_{op}=T_L^{-1}\int _0^{T_L} V_{op}dt $,
with $T_L=2\pi /\omega _L$. The quadratic dependence on $V$ ensures
that the optical potential does not average to zero.

The first type of contribution in $V$ is dominant for dipole transitions and
gives the usual dipole term
\begin{equation}
\tilde{V}_{op}^{D}(X)= \frac{\hbar |\Omega _{D}|^2}{4\Delta } \cos ^2 (k_L X)
\end{equation}
with the Rabi frequency $\Omega _{D}=(eA_0/\hbar cm)<e|\hat{p}_z|g>$.
Similarly, for quadrupole transitions the potential reduces to the quadrupole
term
\begin{equation}
\tilde{V}_{op}^ {Q}(X)= \frac{\hbar |\Omega _{Q}|^2}{4\Delta } \sin ^2 (
k_L X)
\end{equation}
where now the Rabi frequency is $\Omega _{Q}=(e A_0k_L
/\hbar cm)<e|x\hat{p}_z|g>$. Note the additional presence of the
factor $k_L$ with respect to the dipole case. The explicit
evaluation of the matrix elements for some atoms can be found in
\cite{Moi}.

In summary, we can write the potentials generated by the standing wave light as
\begin{equation}
\tilde{V}_{op}^{D}(X) = V_D^0 \cos ^2 (k_LX)
\end{equation}
when the light grating frequency is close to a dipole transition, whereas for a
quadrupole one we have
\begin{equation}
\tilde{V}_{op}^{Q}(X) = V_Q^0 \sin ^2 (k_LX)
\end{equation}
As signaled before, the intensity of the potentials, $V_D^0$ and $V_Q^0$, is
determined by the Rabi frequencies.

\section{Diffraction patterns}

The knowledge of the interaction potential allows one to evaluate
the diffraction patterns. We shall consider two situations. The
first one is the interaction of the atomic beam with a laser only
having one mode, close to the quadrupole
transition. The second one corresponds to a laser with two modes,
one close to the dipole transition and the other to
the quadrupole one. On the other hand, as it is well-known, there
are two regimes in the Kapitza-Dirac effect, diffraction for
thin light gratings and Bragg's scattering for thick ones. Then for
each situation we must consider separately the two regimes.

\subsection{Patterns for a single-mode laser}

We consider first diffraction. We assume that the momentum of the
atoms is large compared to that of the photons. Then the kinetic
energy remains approximately constant and may be neglected. We can
use the Raman-Nath approximation. In addition to the Raman-Nath
approximation two conditions must be fulfilled to derive the
diffraction pattern in the usual way \cite{Ada}. On the one side,
the rate of spontaneous emission must be small during the
interaction. Mathematically, this condition reads $\Delta \gg
\Gamma$, where $\Delta $ is the detuning and $\Gamma $ the decaying
rate. On the other hand, the evolution must be adiabatic, $\Delta
> 1/\tau $, with $\tau $ the interaction time.

If the atom is initially in the state $\exp (ik_0 X)$, the final one is
\begin{equation}
e^{i\tilde{V}_{op}^{Q}(X)\tau /\hbar} e^{ik_0X}=e^{iV_{Q}^0 \tau /2\hbar} \sum
_{n=-\infty}^{\infty } i^n J_n \left( \frac{V_{Q}^0 \tau}{2\hbar } \right)
e^{i(2nk_L +k_o)X}
\end{equation}
where we have used the formula $\exp (i\xi \cos \varphi )= \sum
_{n=-\infty }^{\infty }i^n J_n (\xi )\exp (in\varphi )$ with $J_n$
the n-th order Bessel function.

Thus, up to irrelevant global phases, we have the same pattern
obtained for dipole-type transitions. In an intuitive picture, the
contributions to each peak correspond to even multi-photon
interchange processes. The first one, $n=1$, represents one-photon
absorption followed by one stimulated emission at the same frequency ($k_0
\rightarrow k_0 +2k_L$). Of course, $n=0$, corresponds to no photon
interchange. The only important difference is that the Bessel
functions depend on $V_Q^0$ instead of $V_D^0$.

Next, we consider if these patterns can be studied experimentally. We
give an approximate evaluation of their intensity, based on the
values presented in \cite{Moi}. Typical values of $V \tau /\hbar$
for the observation of Kapitza-Dirac diffraction are of the order of
unity \cite{Bat}. Some of the experiments demonstrating the effect
have been carried out with Na atoms with the values $\tau \approx
(1/14) \times 10^{-6} s$ and $V/\hbar \approx 18 \times 10^6
s^{-1}$, that is, we must consider potentials of the order of $V
\approx 10^{-8}eV$. On the other hand, for the quadrupole-type decaying rate we
have $\Gamma \approx 10^3 s^{-1}$; using a detuning $\Delta \approx 1
GHz$ (the detuning used in \cite{Moi} in the numerical simulations) we guarantee
adiabatic evolution and the absence of spontaneous emission. The relation
between the maximum potential induced by a light grating in a frequency close to
the quadrupole transition for Na atoms and the laser intensity has been
evaluated in \cite{Moi}. Using these data we have that the above values of the
potential can be reached with an intensity of $10^8$ to $10^9 W/m^2$, not too
different from those used in the dipole case. The same experimental
procedure used in that case seems to be adequate for the
verification of our proposal.

In the Ca case a potential of $10^{-8}eV$ can be reached with
intensities close to $10^8 W/m^2$, for a detuning of $70 kHz$
\cite{Moi}. Taking as in the Na case an interaction time of $10^{-7}
s$ we need, in order to have an adiabatic evolution, a detuning of
$\Delta \approx 10^8 s^{-1}$. Note that with this choice the rate of
spontaneous emission is negligible, because for the Ca case we have
$\Gamma \approx 300 s^{-1}$. Assuming a relation of the type $V\sim
I/\Delta $ (as it is true in the perturbative case \cite{Moi}), we
would need an intensity $I \approx 10^{11}W/m^2$ for $\Delta \approx
10^8 s^{-1}$, that is, four orders of magnitude above the usual
values in this type of experiments. For these intensities, one
generates the standing wave using counter-propagating laser pulses,
whose duration is only about $10 ns$ \cite{Ba2}, shorter than the
interaction time, making useless the usual schemes for our system.
We conclude that the case of Na atoms can be tested with the
standard techniques, being the Ca case much more demanding.

In Bragg's scattering, a simple calculation (see, for instance
\cite{Bat}, replacing the $\cos$-type function by a $\sin$-type one)
shows that the transmission and scattering amplitudes are the same
of the dipole approximation with the change of $V_D^0$ by $V_Q^0$.
In particular, the transmission and scattering probabilities are
\begin{equation}
P_{tra}=\cos ^2 \left( \frac{V_Q^0 \tau}{4\hbar } \right)  \; ; \; P_{sca}=\sin
^2 \left( \frac{V_Q^0 \tau}{4\hbar }  \right)
\end{equation}
For real experiments with Na (and dipole transitions) the
interaction times are of the order of $10^{-5}s$ and the potentials
of $10^{-10}eV$ \cite{Bat}. The detuning is again of the order of $1
GHz$. For the Ca and Na atoms these values can be reached with laser
intensities of, respectively, $10^8$ and $10^7 W/m^2$ \cite{Moi}.
Although these intensities are lower than in the case of
diffraction, the demonstration of Bragg's scattering is in general
more demanding, because now the focusing of the laser and the
coherence properties of thick standing waves are more difficult to
implement \cite{Ba3}.

\subsection{Patterns for a two-mode laser}

Now, we consider a two-mode laser. One of the modes has a frequency
close to the quadrupole transition, and the other to the dipole one.
We consider, for simplicity, the case of the Na atom which, as seen
before, seems to be accessible to the experimental scrutiny.

First of all, we need the potential for the light-atom interaction. In the case
of two laser modes inducing dipole transitions the potential can be
approximately expressed as the sum of the two optical potentials associated with
the modes (see, for instance, a short discussion in \cite{Ros}). In the case of
two modes related to dipole and quadrupole transitions we expect a similar
behavior, specially taking into account that the crossed effects between the two
transitions can be neglected. Then we can use a two-mode light potential of the type
\begin{equation}
V_{TM} \approx V_D^0 \cos ^2 (k_Dx) + V_Q^0 \sin ^2 (k_Q x)
\end{equation}

In general, $V_D^0$ and $V_Q^0$ can be different, although we choose
them of the same order of magnitude. We denote by $k_D$ and $k_Q$
the wave vectors of the laser modes associated with the dipole and
quadrupole transitions. The wave function after the interaction is
\begin{eqnarray}
e^{iV_{TM}(X)\tau /\hbar} e^{ik_0X}= e^{iV_Q^0\tau /2\hbar} e^{iV_D^0\tau /2\hbar}
\sum _{n=-\infty}^{\infty} \sum _{m=-\infty}^{\infty} i^{n+m} \times \nonumber
\\
J_n \left( \frac{V_D^0\tau}{2\hbar} \right) J_m \left(
-\frac{V_Q^0\tau}{2\hbar} \right) e^{i(2nk_D+2mk_Q+k_0)X}
\end{eqnarray}
\begin{figure}[H]
\center
\includegraphics[width=10cm,height=8cm]{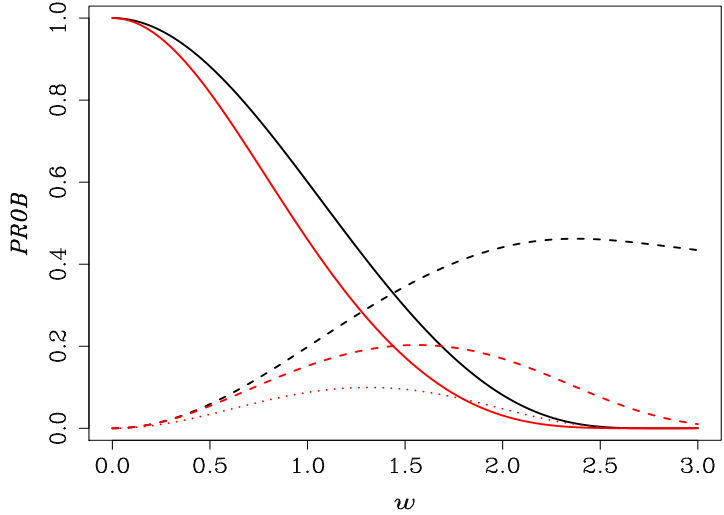}
\caption{Probability of detection of atoms in the lower diffraction
orders versus the dimensionless parameter $w=V_D^0\tau /\hbar$. The
black and red curves represent respectively the single- and two-mode
cases. The continuous, dashed and dotted curves correspond to the
values $n=0$ (and $m=0$ for two modes), $n=1$ (and $m=0$ for two
modes), and $n=0$ and $m=1$ (only for two modes). We use for the
intensity of the potentials the values $V_D^0=1$ and $V_Q^0=0.8$.}
\end{figure}
Several interesting properties easily follow from this expression.
The probability for the central or no-diffraction peak, $n=m=0$, is
$J_0^2(V_D^0\tau /2\hbar)J_0^2(V_Q^0\tau /2\hbar)$ that differs from
the single-mode probability for this case. Having two channels of
diffraction, the contributions of both must be present. For any
other diffraction order the number of peaks doubles. For instance,
for the first order we have the peaks $n=\pm 1$ ($m=0$) and $m=\pm
1$ ($n=0$), which take place for different momentum interchanges,
$\pm 2\hbar k_D$ and $\pm 2\hbar k_Q$. The amplitude of, for
instance, the peak $n=1$ ($m=0$) is $J_1^2(V_D^0\tau
/2\hbar)J_0^2(V_Q^0\tau /2\hbar)$. It implies again a modification
with respect to the single mode case: the detection probability in
that channel is modulated by the coefficient $J_0^2(V_Q^0\tau
/2\hbar)$, associated with the presence of a second channel.

We represent the amplitudes of the non-diffraction and first
diffraction peaks in Fig. 1. The detection probabilities are always
smaller in the two-mode arrangement.

Finally, we consider Bragg's scattering in the two-mode scenario. In
order to have Bragg's scattering, the modulus of the longitudinal
wave vector of the incident particle must be equal (or very close)
to one of the wave vectors of the laser modes, that is, $k_0=\pm
k_D$ or $k_0=\pm k_Q$. Being $k_D \neq k_Q$ we can only have a type
of Bragg-scattering. At variance with the diffraction regime, we do
not have modifications of the scattering patterns in the two-mode case.

\section{Conclusions}

We have analyzed in this paper the atomic Kapitza-Dirac effect with
quadrupole transitions, instead of the dipole ones usually
considered in the literature. The effect is, in principle,
accessible to experimental scrutiny. For some atoms (for instance,
Na) the diffraction patterns associated with quadrupole atomic
transitions are observable with experimental parameters (laser
intensity,..) close to those associated with dipole transitions.

In addition to the demonstration of the existence of observable
quadrupole effects in this regime of the matter-light interaction
our scheme could be interesting in some other aspects. Our proposal
would be the first example of a diffraction arrangement driven by a
quadrupole force. From a more practical point of view, the
diffraction of atoms can be used to measure the transition
quadrupole matrix elements, by fitting the experimental data to the
distributions $|J_n(V_Q^0\tau /2\hbar)|^2$. The values of the matrix
elements are of interest to validate calculational methods.

{\bf Acknowledgments} I acknowledge support from Spanish Ministerio de
Ciencia e Innovaci\'on through the research project FIS2009-09522.

\end{document}